\documentstyle[prl,aps,multicol,tighten,epsf]{revtex}

\begin{document}
\draft

\title{Universal algebraic relaxation of fronts propagating into
an unstable state}
\author{Ute Ebert and Wim van Saarloos}
\address{Instituut--Lorentz, Universiteit Leiden, Postbus 9506, 
2300 RA Leiden, the Netherlands}
\date{July 15, 1997}
\maketitle

\begin{abstract}
We analyze ``pulled'' or ``linearly marginally stable'' fronts 
propagating into unstable states. While ``pushed'' 
fronts into meta- and unstable states relax exponentially, pulled 
fronts relax algebraically, and simultaneously the standard 
derivation of effective interface equations breaks down. 
We calculate all universal relaxation terms of uniformly translating
pulled fronts. The leading $1/t$ and $1/t^{3/2}$ corrections to the
velocity are determined by the dispersion relation 
of the linearized equation only. Our analysis sheds new light 
on the propagation mechanism of pulled fronts.
\end{abstract}


\begin{multicols}{2}

Consider systems far from equilibrium with a continuous order 
parameter, where a stable state propagates into an unstable state, 
and assume, that thermal perturbations can be neglected.  
Some experimental examples we discuss below. If the initial profile 
is steep enough, arising, e.g., through a local initial perturbation, 
it is known \cite{aw,bj,stokes,vs2,oono} that the propagating front 
in practice always relaxes to a unique profile and velocity. 
Depending on the nonlinearities, one can distinguish
two regimes: as a rule, fronts whose propagation is driven (``pushed'')
by the nonlinearities, resemble very much fronts propagating into
metastable states and their relaxation is exponential in time.
This regime is often refered to as ``pushed'' \cite{stokes,oono} or
``nonlinear marginal stability'' \cite{vs2}. If, on the other hand, 
nonlinearities mainly cause saturation, fronts propagate with a
velocity determined by linearization about the unstable state,
as if they are ``pulled'' by the linear instability (``pulled''
or ``linear marginal stability'' regime). In the simplest 
case [Eq.\ (\ref{1})] of a pulled front it has been proven rigorously 
\cite{bramson} that the velocity relaxation is proportional to $1/t$ 
in leading order, and there are some heuristic arguments \cite{vs2},
that this is generally true for pulled fronts. 
In this paper, we identify the general mechanism leading to
slow relaxation of uniformly translating fronts and use it to
introduce a systematic analysis which allows us to determine
{\em all universal asymptotic terms}. We get a unification 
and extension of various seemingly unrelated approaches 
\cite{vs2,oono,bramson,landau,powell} as a bonus.

Our present investigation actually was motivated by an intimately 
related problem, namely the derivation of effective interface 
equations for some order parameter field varying on a short length 
scale and coupled to some external field (like temperature
in a solidification or combustion front) varying on some outer 
length scale \cite{fife}. In such problems, it is well known that 
in the limit in which the curvature of the front or domain wall is 
much smaller than its width, smooth interface problems reduce to 
those with a sharp interface with boundary conditions that are 
local in space and time. This connection actually lies at the basis 
of recent advances in the numerical studies of interfacial growth 
problems like dendrites \cite{karma}. Although this is often not 
made explicit, this reduction of a smooth front problem to a sharp 
interface formulation amounts to an adiabatic decoupling of the
interface relaxation from the dynamics of the outer field(s). 
This decoupling is only possible if the profile of the front 
propagating into a metastable state relaxes {\em exponentially
fast} to its asymptotic shape and speed. If the order parameter
front is a pulled front, the standard derivation of effective 
interfaces \cite{fife} breaks down and at the same time the 
relaxation becomes non-exponential. Both are due to the properties 
of the spectrum of the operator ${\cal L}^*$ derived by linearizing 
about the asymptotic front, as we will discuss in more detail 
elsewhere \cite{ebert2}. 

We here analyze a pulled uniformly translating front in one dimension 
without external fields and its algebraic relaxation towards its 
asymptotic shape. We find that not only every sufficiently steep initial 
profile relaxes to a unique asymptotic front profile, but that also this 
profile is asymptotically approached along a unique trajectory.
Such properties, that are independent of the precise initial
conditions, we call universal. We construct all universal terms 
explicitly in an asymptotic $1/\sqrt{t}$ expansion and confirm
our predictions numerically. We not only calculate the dominant $1/t$ 
term of the velocity \cite{bramson,vs2}, but also find the subdominant
$1/t^{3/2}$ term. The term $1/t^2$ is nonuniversal, since it at least
partially can be generated from the $1/t$ term through a temporal 
shift of the initial conditions $t \to t+t_0$. Furthermore we, for 
the first time, predict the relaxation of the shape of the profile. 
Technically, we here construct the contracting center manifold in 
function space about the asymptotic front profile.

We use the word ``universality'' in analogy to scaling in 
critical phenomena: Not only do a large class of initial 
conditions reach a unique asymptotic profile (fixed point) in an 
infinite-dimensional function space, but also the asymptotic 
approach towards this fixed point is completely determined by 
the expansion about it. This is reminiscent of the universal 
corrections to scaling.

Fronts propagating into an unstable state arise in various fields of
physics: they are important in many convective instabilities in
fluid dynamics such as the onset of von Karman vortex generation
\cite{provansal}, in Taylor \cite{ahlers} and Rayleigh-B\'enard
\cite{fineberg} convection, they play a role in spinodal
decomposition near a wall \cite{ball}, the pearling instability of
laser-tweezed membranes \cite{goldstein}, the formation of kinetic,
transient microstructures in structural phase transitions
\cite{salje}, the propagation of a superconducting front into an
unstable normal metal \cite{dorsey}, or in  error propagation in extended 
chaotic systems \cite{torcini}. Our own interest in the relaxation of 
pulled fronts stems from our search for an interfacial description for 
so-called streamers, which are dielectric breakdown fronts \cite{ebert}.
The experimental relevance of front relaxation is illustrated on 
propagating Taylor vortex fronts. Here the measured velocities were 
about 40\% lower than predicted theoretically, and only later numerical 
simulations \cite{luecke} showed that this was due to slow transients.

Our analysis can be formulated quite generally for partial
differential equations which are of first order in time but of 
arbitrary order in space, as long as they admit uniformly translating 
``pulled'' solutions. (``Pulling'' is explained in the next paragraph.) 
For ease of presentation we guide our discussion along two examples which 
we have investigated analytically as well as numerically.
\begin{equation}
\label{1}
\partial_t \phi(x,t) = \partial_x^2 \phi + f(\phi)~,~~~
f(\phi)=\phi-\phi^3~,
\end{equation}
with $\phi$, $x$, $t$ real, is the nonlinear diffusion equation. 
It is a prototype equation for pulled fronts. It is also known 
as KPP equation (after Kolmogorov {\it et al.}), Fisher 
equation, or FK equation. In (\ref{1}), the state $\phi=0$ is unstable 
and the states $\phi=\pm 1$ are stable. We consider a situation where 
initially $\phi(x,0)$ asymptotically decays quicker than $e^{-x}$ for 
large $x$, or in particular one with $\phi(x,0)\ne0$ in a localized 
region only. The region with $\phi\ne0$ expands in time, and a propagating 
front evolves. It has been proven rigorously, that relaxation is always 
to a unique front profile $\phi^*(x-v^*t)$ with velocity $v^*=2$ \cite{aw}, 
and that the velocity relaxes asymptotically as $v(t)=2-3/(2t)$
\cite{bramson}. The extension
\begin{equation}
\label{2}
\partial_t \phi(x,t) = \partial_x^2 \phi - \gamma \partial_x^4 \phi
+ \phi - \phi^3~
\end{equation}
is often refered to as the EFK (extended FK) equation, and serves 
as a model equation for higher order equations that admit front 
solutions. The rigorous methods of \cite{aw,bramson} are not 
applicable here, but the algebraic relaxation behavior towards a 
unique uniformly translating front with velocity $v^*$ continues 
for $0\le\gamma<1/12$ \cite{dee,vs2}.

Since the basic state $\phi=0$ into which the front propagates, is
linearly unstable, even a small perturbation around $\phi=0$ grows 
and spreads by itself. According to the {\em linearized} equations 
any localized small perturbation will spread asymptotically for
large times with the ``linear marginal stability'' speed $v^*$ 
\cite{landau}
\begin{equation} \label{v*}
\left. { {\partial \Im \omega}\over{\partial \Im k}}\right|_{k^*} - v^*
  =0~,~~
\left.  {{\partial \Im \omega}\over{\partial \Re k}}\right|_{k^*} =0~,~~
{{\Im \omega (k^*)}\over{\Im k^*}}=v^*~,
\end{equation}
where $\omega(k)$ is the dispersion relation of a Fourier mode
$e^{-i\omega t+ikx}$.
The first two equations in (\ref{v*}) are saddle point equations
in the complex $k$ plane that govern the long time asymptotics 
of the Green's function in a frame moving with the leading edge of 
the front. The third equation expresses that for selfconsistency, 
the linear part of the front should neither grow nor decay in the 
comoving frame.
If the nonlinearity pushes the front to a velocity $v^\dag>v^*$, 
we call the front nonlinearly marginally stable or pushed.
If the front propagates with velocity $v^*$, we call 
the front linearly marginally stable or pulled.
Replacing the nonlinearity in (\ref{1}) by, e.g., 
$f(\phi)=\epsilon\phi+\phi^3-\phi^5$, one finds
the fronts to be pushed for $\epsilon<3/4$, and to be pulled
for $\epsilon>3/4$ \cite{vs2}.
We stress that for a given partial differential equation, the
pulled speed $v^*$ and the asymptotic shape of the profile are
determined explicitly by the dispersion relation \cite{bj,vs2}. 

Our analysis and predictions can now be summarized as follows: 
for Eqs.\ (\ref{1}) and (\ref{2}) (with $0\le\gamma<1/12$), and others 
where localized initial conditions develop into uniformly traveling 
pulled fronts, i.e., into fronts with asymptotic speed $v^*$ and 
$\Re k^*=0=\Re\omega(k^*)$, the asymptotic relaxation for $t\to\infty$ 
is given by
\begin{eqnarray}
\label{phi}
\phi(\xi,t) &=& \phi^* (\xi)  + \eta(\xi,t)~,~~~ \xi=x-v^*t -X(t)~,
\\
\label{xdot}
\dot{X}(t)&=&\frac{-3}{2\Lambda t}
\left(1-\frac{\sqrt{\pi}}{\Lambda \sqrt{Dt}}\right) + 
O\left(\frac{1}{t^2}\right)~.
\end{eqnarray}
Here $D= \frac{1}{2} \partial^2 \Im\omega/(\partial \Im k)^2|_{k^*}$
plays the role of a diffusion coefficient for perturbations about 
the asymptotic shape, and $\Lambda=\Im k^*>0$ is the asymptotic decay rate
of $\phi^*$: $\phi^*(\xi)\propto (\xi+const.) \;e^{-\Lambda\xi}$
for $\xi\to\infty$. The factor $(\xi+const.)$ is due to two roots 
of $\omega(k)-vk$ coinciding, which is implied by the saddle point 
equations (\ref{v*}). Note that all quantities in (\ref{xdot}) are 
determined by the linear dispersion relation $\omega(k)$. 

It is central to the analysis, which we will further elucidate below, 
that the coordinate system, in which the asymptotic shape $\phi^*$ is 
subtracted, moves with the speed $v^*+\dot{X}$ of $\phi(\xi,t)$. 
Only then the shape correction $\eta$ stays small for all times. 
We then find through some expansion in the ``interior region'' 
of the front, where $|\eta|\ll\phi^*$,
\begin{eqnarray}
\label{shape}
\phi(\xi,t) &=& \phi_{v^*+\dot{X}}(\xi) + O(t^{-2}) 
\\
&=& \phi^*(\xi) + \dot{X} \eta_{sh}(\xi) + O(t^{-2})
~,~\eta_{sh}=(\delta\phi_v/\delta v) \Big|_{v^*}~.
\nonumber
\end{eqnarray}
In the far edge, where $\xi \gtrsim O( \sqrt{Dt}) \gg 1$, a different 
expansion is needed, as the transient profile $\phi$ falls off
faster than $\phi^*$, so that $\eta \approx -\phi^*$.
Matching to the interior (\ref{shape}) and imposing that
the asymptotic shape $\phi^*$ is approached for $t\to\infty$,
and that the transients are steeper than $e^{-\Lambda\xi}$ for 
$\xi\to\infty$, uniquely determines the velocity correction 
$\dot{X}$ (\ref{xdot}) and the intermediate asymptotics
\begin{equation}
\label{faredge}
\phi(\xi,t)  \approx e^{-\Lambda \xi
  -\xi^2/(4Dt)} ( \xi + const. + O(1/\sqrt{t}))~.
\end{equation}

Let us now first discuss three physical observations, that are crucial
ingredients of our systematic analysis. We then present our numerical 
results and finally briefly sketch our calculation.

$(i)$ The leading $1/t$ term in $\dot{X}(t)$ in (\ref{xdot}) can be 
understood intuitively through a heuristic argument
\cite{vs2}: If we start from localized initial conditions 
and analyze in the asymptotic frame $\xi^*\!=\!x\!-\!v^*t$,
$\phi(\xi^*,t)$ should approach $\phi^*(\xi^*)$ as $t\to\infty$, 
but for a fixed time, $\phi$ should fall off faster than $\phi^*$ 
as $\xi^* \to\infty$. To study this crossover, consider for simplicity Eq.\
(\ref{1}); if we linearize, and substitute $\phi(\xi^*,t)\!=\!e^{-\Lambda
  \xi^*}\psi(\xi^*,t)$ (with $v^*\!=\!2$, $\Lambda\!=\!1\!=D$ in this case), 
we get the simple diffusion equation $\partial_t \psi \!=\!
\partial^2_{\xi^*} \psi$. Clearly, the similarity solution which
matches to $\phi^*(\xi^*) \!\sim\! e^{-\Lambda\xi^*}(\xi^* \!+\!
\mbox{const.})$ is $\psi \sim (\xi^*/t^{3/2}) e^{-{\xi^*}^2/4t}$, so
$\phi\! \sim \!e^{[-{\Lambda}\xi^* -3/2 \ln t \!+\! \ln \xi^* -
  {\xi^*}^2/4t]}$ \cite{note3}. Hence, if we now track the position
$\xi^*_h$ of constant height $h\!\ll\! 1$, which is defined as
$\phi(\xi_h^*,t)\!=\!h$, we find $\xi^*_h (t)\!=\!
-3/(2\Lambda)\ln t +\ldots$ in the frame $\xi^*$. 
This is precisely the leading term of $X(t)$!

$(ii)$ Observation $(i)$ shows that within the frame 
$\xi^*=x-v^*t$, the leading edge of the profile
moves back a distance $-3/(2\Lambda)\ln t$. But this {\em must be true 
for all heights $h$}, not just for $h \ll 1$: The front width $W$ is 
finite in equations like (\ref{1}) and (\ref{2}). So positions 
$\xi^*_h(t)$ are related through $\xi^*_{h'}(t)-\xi_h^*(t)=O(W)=O(1)$. 
In other words, the dominant shift $-3/(2\Lambda) \ln t$ determined 
from the leading edge $h\ll1$, is the same for all amplitudes $h$.
This is why in order to study the asymptotics, we {\em have to} 
perturb about the asymptotic shape $\phi^*(\xi)$ and not about 
$\phi^*(\xi^*)$, since the two profiles in frames $\xi$ or $\xi^*$
are pulled arbitrarily far apart due to the diverging logarithmic 
shift $X(t)$. A very unusual situation indeed!

$(iii)$ In (\ref{shape}), $\eta_{sh}$ is the shape mode $\delta
\phi_v/\delta v|_{v^*}$ which gives, to linear order, the change in
the shape of the uniformly translating profile $\phi_v$ when varying 
$v$ about $v^*$ \cite{note1}. Since the instantaneous velocity of the 
profile position is just $v^*+\dot{X}(t)$ ($<v^*$), (\ref{shape})
expresses that up to order $t^{-2}$ the front profile in the interior 
front region is just the uniformly translating profile $\phi_v$ with 
velocity $v=v^*+\dot{X}(t)$. This was actually conjectured for Eq.\
(\ref{1}) by Powell {\em et al.} \cite{powell}, based on numerical 
observations, and our analysis gives its first derivation 
in leading order; in order $1/t^2$ the conjecture does not hold. 

We have tested our predictions by numerically integrating
Eqs.\ (\ref{1}) and (\ref{2}) forward in time, starting from localized
initial conditions. In Fig.\ 1{\em (a)} and {\em (b)}, we present
velocities $v_h(t)$ of various amplitudes $h$, in {\em (a)} for 
Eq.\ (\ref{1}), and in {\em (b)} for Eq.\ (\ref{2}) with $\gamma=0.08$. 
Note that the critical value of $\gamma$ is $\gamma_c=1/12=0.083$. 
Eqs.\ (\ref{phi})-(\ref{shape}) imply that in the lab frame $x$,
$v_h(t)=v^*+\dot{X}(t)+g(h)/t^2$
(where $g(h)$ can be expressed in terms 
of $\eta_{sh}$ and $\partial_\xi \phi^*$). Thus we plot $v_h(t)\!-\!v^*\!-\!\dot{X}(t)$ versus $1/t^2$ for various $h$.
According to our prediction, all curves should then converge linearly 
to zero as $1/t^2\to 0$. Clearly, the numerical simulations fully 
confirm this for both equations. 

For our prediction (\ref{shape}) of the shape relaxation, the most 
direct test is to plot 
$[\phi(\xi,t)-\phi^*(\xi)]/(\dot{X}(t)  \eta_{sh}(\xi))$ 
as a function of $\xi$ for various times. This ratio should converge 
to 1 for large times. As Fig.\ 1{\em (c)} shows, this is fully borne 
out by our simulations of the nonlinear diffusion equation (\ref{1}).
Moreover, the crossover for large positive $\xi$ is fully in accord 
with our result that the proper similarity variable in the far edge 
is $\xi^2/t$ --- see Eq.\ (\ref{faredge}). 

We finally give a brief sketch of the systematic analysis, taking the
nonlinear diffusion equation (\ref{1}) as an example. Full details 
will be published elsewhere \cite{ebert2}.

We first consider the ``front interior'' region, where the deviation
$\eta(\xi,t)$ of $\phi$ about  $\phi^*(\xi)$ is small, i.e., $|\eta| \ll
\phi^*$. As there is some freedom   
in choosing $\xi$ due to translation invariance, we choose quite 
arbitrarily the condition, that $\phi(0,t)=\frac{1}{2}=\phi^*(0)$, 
so that $\eta(0,t)=0$, as was also done in Fig.\ 1{\em (c)}. 
Substituting (\ref{phi}) into (\ref{1}), we obtain
\begin{eqnarray}
\label{5}
\partial_t \eta &=& {\cal L}^* \eta + \dot{X}\;
\partial_\xi(\eta\!+\!\phi^*) + \frac{f''(\phi^*)}{2}\;\eta^2 + 
O(\eta^3)~,
\\
\label{6}
{\cal L}^* &=& \partial_\xi^2 + v^* \partial_\xi +f'(\phi^*(\xi)) ~.
\end{eqnarray}
The inhomogeneity $\dot{X}\partial_\xi\phi^*$ in (\ref{5}) is due to
the fact that $\phi^*(\xi)$ is a solution of (\ref{1}) only if $\dot{X}=0$. 
Since $\dot{X}(t)=O(t^{-1})$, and since in the front
interior $|\eta|\ll\phi^*$, the inhomogeneity induces an ordering
in powers of $1/t$, which suggests an asymptotic expansion as
\begin{eqnarray}
\label{7}
\dot{X} & = &\frac{c_1}{t}+\frac{c_{3/2}}{t^{3/2}}+\frac{c_2}{t^2}+\ldots~,
\\
\label{8}
\eta(\xi,t)&  = & \frac{\eta_{1}}{t}+
\frac{\eta_{3/2}}{t^{3/2}}+\ldots~.
\end{eqnarray}
The necessity for actually expanding in powers of $1/\sqrt{t}$ 
emerges  from matching 
to the similarity solutions in the far edge. 
Substitution of the above expansions in (\ref{5}) yields a hierarchy 
of {\em o.d.e.}'s of second order
\begin{eqnarray}
\label{9} &&
{\cal L}^* \eta_1 = - c_1\partial_\xi \phi^* 
~~,~~
{\cal L}^* \eta_{3/2} = - c_{3/2}\partial_\xi \phi^*
\\
\lefteqn{
{\cal L}^* \eta_2 = - c_2\partial_\xi \phi^* 
- c_1\partial_\xi \eta_1 -\eta_1 - f''(\phi^*) \eta_1^2/2 ~~\mbox{etc.}}
\nonumber
\end{eqnarray}
The hierarchy is such that the equations can be solved order
by order. Each $\eta_i$ is uniquely determined by its differential 
equation, the appropriate boundary conditions and the requirement
$\eta_i(0)=0$. The equations for $\eta_1/c_1$ resp.\ 
$\eta_{3/2}/c_{3/2}$ are precisely the differential equation
for $\eta_{sh}=\delta\phi_v/\delta v|_{v^*}$, cf.\ Eqs.\
(\ref{phi}), (\ref{shape}) and paragraph $(iii)$. 

By expanding the $\eta_i$ for large $\xi$, one finds that they all
behave like $e^{-\Lambda\xi}= e^{-\xi}$ times a polynomial in $\xi$,
whose degree grows with $i$.
The $\eta_i$ expansion is therefore not properly ordered for large
$\xi$. This just reflects the fact that on the far right, $\eta$
and $\phi^*$ must almost cancel each other. This is required for fronts
that emerge from localized initial conditions, whose total profile 
thus decays faster than $\phi^*$. A detailed investigation of
this region shows that $z=\xi^2/4t$ is a proper similarity variable
here, and suggests that here the proper expansion is
\begin{equation}
\label{rightexp}
\phi(\xi,t) = e^{-\xi -z} \left[
\sqrt{t}\; g_\frac{-1}{\;2}(z) + g_0(z) + 
\frac{g_\frac{1}{2}(z)}{\sqrt{t}}+ ..
\right]~.
\end{equation}
Upon substitution of this expansion into the original partial
differential equation, linearized about $\phi=0$, we now find a
different hierarchy of ordinary differential equations for the
functions $g_{n/2}(z)$. In this case, the conditions to be imposed on
the $g_{n/2}$'s is that they do not diverge as $e^z$ as
$z\!\rightarrow\!\infty$, and that they match, in the language of
matched asymptotic expansions, the large $\xi$ ``outer''
expansion of the ``inner'' solution based on the $\eta_i$ \cite{note2}. 
These conditions fix the parameters $c_1$ and 
$c_{3/2}$ in (\ref{7}), and this yields the solution given in Eqs.\
(\ref{phi})-(\ref{faredge}) \cite{ebert2}. The structure of the
analysis is essentially the same for higher order equations like (\ref{2}).

In summary, our results show that the $1/t$ relaxation of pulled
fronts is essentially due to the crossover to a Gaussian shaped tip
in the leading edge of the front. The nonlinearities dictate the 
asymptotic tip shape $\phi^*\propto \xi e^{-\Lambda\xi}$ for $t\to\infty$
and $\xi$ large. This asymptote determines the coefficients and 
the $1/t^{3/2}$ term in the velocity correction $\dot{X}$ (\ref{xdot}).
We finally note that analytical arguments as well as numerical
simulations indicate that many of the above arguments can be
generalized to the case of pattern forming fronts, occuring, e.g., 
in Eq.\ (\ref{2}) for $\gamma> 1/12$ or in the Swift-Hohenberg
equation \cite{vs2}. Work is in progress.

This work was started in collaboration with C.\ Caroli and we thank
her for helpful discussions. The work of UE is supported by the 
Dutch Science Foundation NWO and EU-TMR network ``Patterns, Noise and Chaos''.

\end{multicols}

\begin{figure}[h]
\setlength{\unitlength}{1cm}
\begin{picture}(18,4)
\epsfxsize=4.8cm
\put(0.2,0){\epsffile{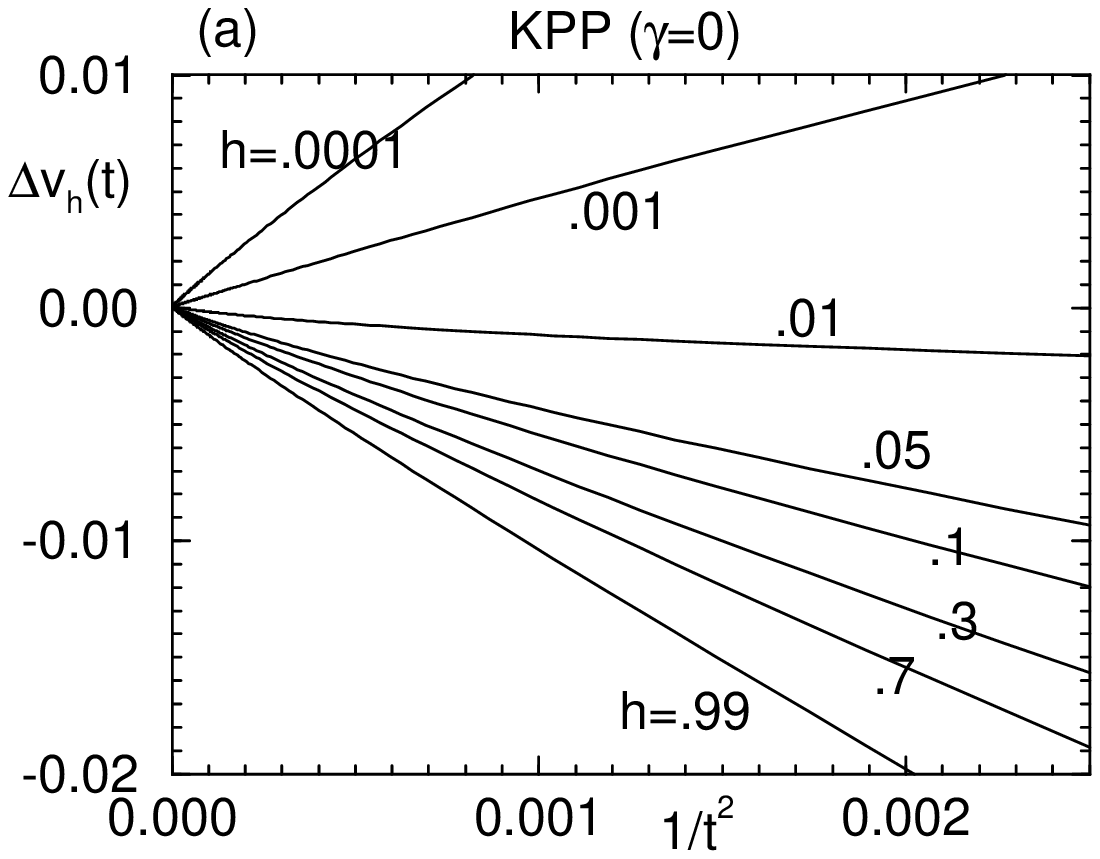}}
\epsfxsize=4.8cm
\put(6.3,0){\epsffile{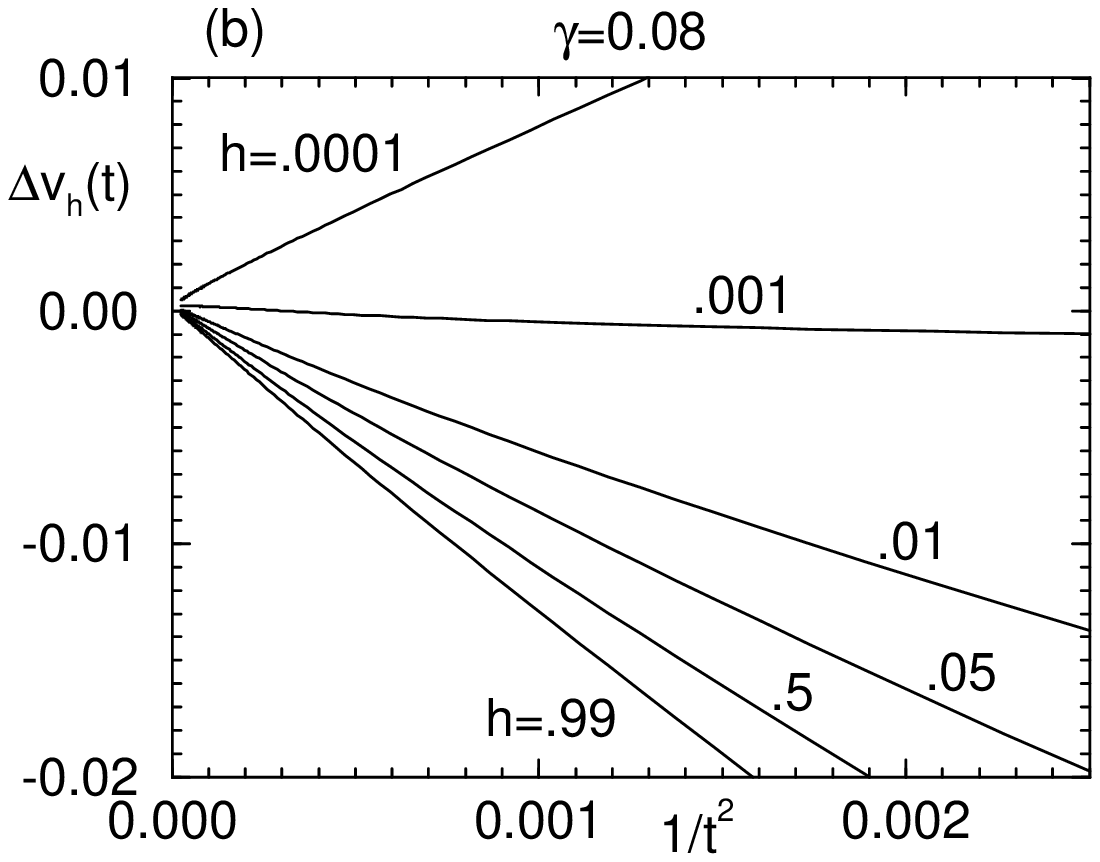}}
\epsfxsize=4.8cm
\put(12.2,0){\epsffile{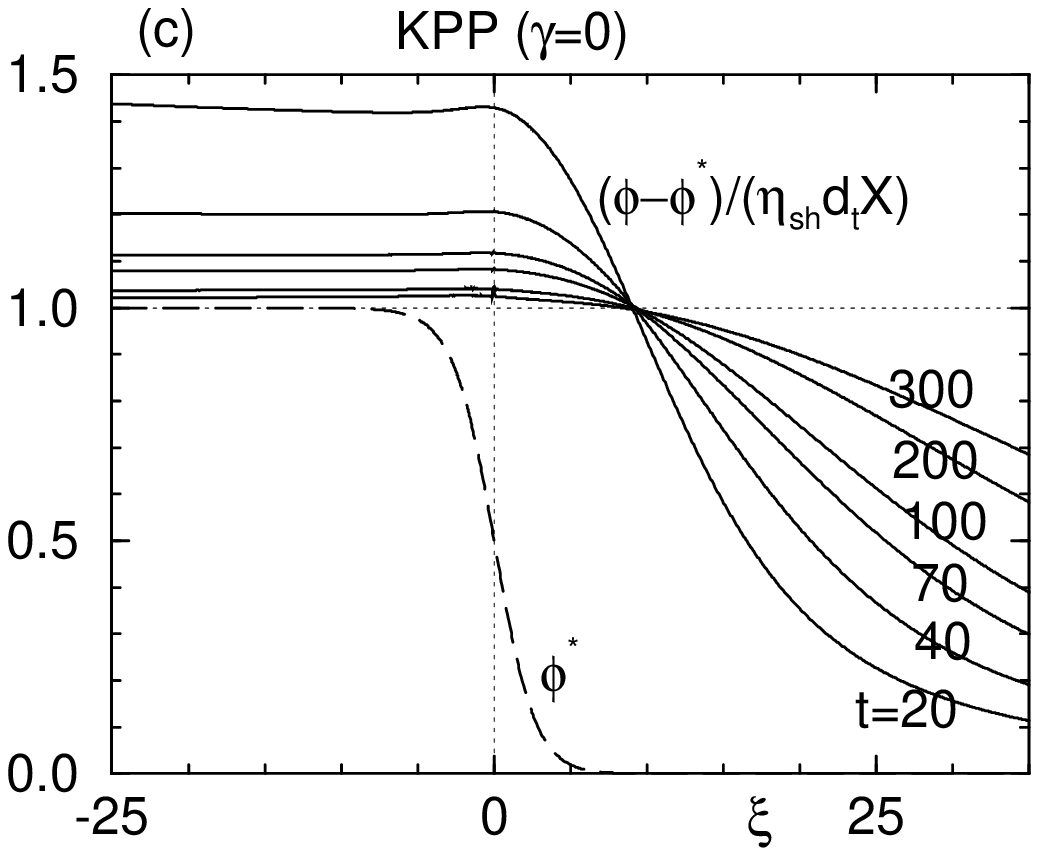}}
\end{picture}
\end{figure}
\small FIG.\ 1. 
{\em (a)} and {\em (b)}: Velocity correction
$\Delta v_h(t) = v_h(t) -v^*-\dot{X}$
as a function of $1/t^2$ for various amplitudes $\phi(x_h,t)=h$,
$v_h=\dot{x}_h$ and for $t \ge 20$. 
{\em (a)}: Eq.\ (1), thus $\gamma=0$, $\Lambda=1=D$, $v^*=2$.
{\em (b)}: Eq.\ (2) with $\gamma=0.08$, thus $D = 0.2$, 
$\Lambda = 1.29$, $v^* = 1.89$.
{\em (c)} Data from {\em (a)} plotted as $(\phi(\xi,t)-\phi^*(\xi))/
(\dot{X} \eta_{sh}(\xi))$ over $\xi$ for various $t$. 
$\phi^*(\xi)$ (dashed) for comparison. 


\begin{references}
\bibitem{note1} At this point, the fact that there is a family of
  uniformly translating profiles also for $v\!<\!v^*$ is crucial!
\bibitem{note2} Specifically, the {\em o.d.e.} for the $g_{n/2}$ has a
  regular singular point at $z\!=\!0$, which admits a $\sqrt{z}$ type
  of singular behavior. The small-$z$
  requirement for $g_{-1/2}$ is that it has a $\sqrt{z}$ term only,
  whereas $g_0$ should not have one.
\end{references}
\end{document}